\documentclass[a4paper,twocolumn,prd,showpacs]{revtex4}
\usepackage{graphicx,color}
\usepackage{revsymb}

\begin{document}

\title{$\delta N$ versus covariant perturbative approach to non-Gaussianity outside the horizon\\ in multifield inflation}
\author{Yuki Watanabe}
\email{yuki.watanabe@physik.lmu.de}
\affiliation{Arnold Sommerfeld Center for Theoretical Physics, Ludwig Maximilian University of Munich,
Theresienstrasse 37, 80333 Munich, Germany}
\date{\today}
\begin{abstract}
We compute the super-Hubble evolution of non-Gaussianity of primordial curvature perturbations in two-field inflation models by employing two formalisms: $\delta N$ and covariant formalisms.
Although the two formalisms treat the evolution of fluctuations radically different, we show that the formulas of the $f_{NL}$ parameter agree quantitatively with each other within $1 \% $ accuracy.
We analytically find that the amplitude of $f_{NL}$ decays no faster than $a^{-3}$ as the inflationary trajectory reaches to the adiabatic limit for generic potentials.
\end{abstract}

\eprint{LMU-ASC 47/11}
\maketitle

\section{Introduction}

Cosmological perturbation theory is a powerful tool to analyze cosmological observations. It has been sophisticated over years by confrontation with more and more precise data of the cosmic microwave background anisotropies \cite{Komatsu:2010fb}. 

In the single-field inflationary scenario the primordial curvature
perturbations are generated during inflation and become the seed of the inhomogeneities
seen today. These fluctuations are observed to follow nearly Gaussian
statistics. It has been shown that all single-field inflation models with the canonical kinetic term produce
almost Gaussian signals if the spectrum is nearly scale invariant \cite{Maldacena:2002vr,Acquaviva:2002ud, Creminelli:2004yq,
Seery:2005wm}\footnote{In the following, we denote the local form of non-Gaussianity by $f_{NL}$. Interestingly, the single-field consistency relation of $f_{NL}$ holds also for non-Einstein gravity (see e.g., \cite{Germani:2011ua}). For the definition of $f_{NL}$, see \cite{Komatsu:2001rj}.}.
The primordial curvature perturbations freeze in at the Hubble horizon
exit, and the amplitude is conserved until the subsequent horizon re-entry regardless of the post-inflationary events \cite{Lyth:2004gb}. 
This fact makes single-field inflationary scenarios very powerful and
predictable; a flip side of this is that no one can know what happened after
inflation and before radiation domination by analyzing large-scale
inhomogeneities on the sky.

However, if curvature fluctuations are generated by some other field than the
inflaton, they evolve until the inflationary trajectory reaches to the adiabatic limit (i.e., all entropy modes decay) and can be non-Gaussian \cite{Linde:1996gt,Lyth:2002my,Dvali:2003em}. 
Non-Gaussianities in multi-field inflation models have been studied extensively (see, e.g., \cite{Bartolo:2001cw,Bernardeau:2002jy,Lyth:2005fi,Lyth:2005du, Seery:2005gb,Rigopoulos:2005us,Tzavara:2010ge, Vernizzi:2006ve, Langlois:2008vk, Langlois:2011zz, Yokoyama:2007dw, Suyama:2010uj, Meyers:2010rg,Meyers:2011mm,Peterson:2010mv,Elliston:2011dr,Byrnes:2006vq,Byrnes:2008wi,Byrnes:2009qy,Battefeld:2006sz,Battefeld:2009ym, Cogollo:2008bi, Rodriguez:2008hy,Kim:2010ud}). In there, the key mechanism on producing non-Gaussian adiabatic perturbations always involves the existence of entropy perturbations on large scales.

The $\delta N$ formalism \cite{Sasaki:1995aw,Sasaki:1998ug, Lyth:2004gb, Lyth:2005fi} is quite useful for the studies of non-Gaussianity since the curvature perturbations can be obtained by computing derivatives of the local volume expansion (e-folding number $N$) with respect to the fields from initially flat hypersurface to finally uniform density hypersurface, where only background equations need to be solved.

On the other hand, in conventional perturbative approaches to the spatial curvature, the perturbed Einstein equations and the perturbed Klein-Gordon equations should be coupled and solved together with background equations. 
Based on Hawking's covariant approach to cosmological perturbations \cite{Hawking:1966qi}, a nonlinear and covariant formalism for
multi-scalar fields was developed by \cite{Langlois:2006vv,RenauxPetel:2008gi,Langlois:2010vx} (see also \cite{Rigopoulos:2005us,Tzavara:2010ge} for an equivalent approach) to solve the evolution of
curvature perturbations, and here we shall apply them to non-Gaussianities on large scales in the two-field inflation.
Both formalisms are fully nonlinear and geometrically transparent, but treatments of the evolution of perturbations are radically different. 
Therefore, it would merit further study of non-Gaussianities by directly comparing the two approaches.

In the $\delta N$ formalism, the non-Gaussianity is given by \cite{Lyth:2005fi}
\begin{eqnarray}\label{eq:fnl_deltan}
\frac35 f_{NL} = \frac{\sum_{I,J}N_IN_JN_{IJ}}{2[\sum_{I}N_IN_I]^2},
\end{eqnarray}
while in the covariant formalism, it is given by
\begin{eqnarray}
\frac35f_{NL}^{\rm transfer}=\frac{4\epsilon_*^2\left[{ T}^{(1)}_{\zeta}\right]^2 { T}^{(2)}_{\zeta}}{ \left[1+2\epsilon_*\left({ T}^{(1)}_{\zeta}\right)^2\right]^2},\hspace{2cm}\label{eq:fnl_transfer}\\
\frac35f_{NL}^{\rm horizon} 
= \hspace{6cm}\nonumber\\
\frac{-\left(\epsilon\eta_{ss}\right)_*  \left[{ T}^{(1)}_{\zeta}\right]^2 
+3\sqrt{\epsilon_*/2} \ \eta_{\sigma s*}{ T}^{(1)}_{\zeta} 
+ \left(\epsilon- \eta_{\sigma\sigma}/2\right)_* }
{ \left[1 + 2\epsilon_*\left({ T}^{(1)}_{\zeta}\right)^2 \right]^2},
\label{eq:fnl_horizon}
\end{eqnarray}
where $ f_{NL} = f_{NL}^{\rm horizon} + f_{NL}^{\rm transfer}$. 
Here $N_I \equiv \partial/\partial\varphi^I\int Hdt$ and $N_{IJ} \equiv \partial^2/\partial\varphi^I\partial\varphi^J\int Hdt$ are from background trajectories, while transfer functions $T_{\zeta}^{(1),(2)}$ represent super-horizon generation of curvature perturbations $\zeta$ from entropy ones.
The subscript * denotes a value evaluated at the Hubble-horizon exit, and we have used the standard slow roll parameters, $\epsilon\equiv -\dot{H}/H^2\ ,\eta_{IJ}\equiv V_{,\varphi^I\varphi^J}/V$. 

In this note, we present a detailed numerical comparison between the two formalisms
in case of two-field inflation models. 
We show that Eqs.~(\ref{eq:fnl_deltan}), (\ref{eq:fnl_transfer}) and (\ref{eq:fnl_horizon}) agree quantitatively within 
$1 \%$ accuracy.
For illustrative purposes, we use canonical scalars with potentials used in the literature as examples, but the generalization is straightforward and the same conclusion holds. 
A similar comparison has been done by \cite{Lehners:2008my,Lehners:2009ja} for ekpyrotic models, but they have ignored the contribution from Eq.~(\ref{eq:fnl_horizon}) resulting in the less accurate quantitative agreement.

We also analyze the fate of $f_{NL}$ when the inflationary trajectory reaches to the adiabatic limit for generic potentials. In this case, we find that the amplitude of $f_{NL}$ converges to its asymptotic value no faster than $a^{-3}$ contrary to the recent claim \cite{Meyers:2010rg}.

We employ the natural unit $\hbar=c=1$ and $8\pi G=M_{\rm Pl}^{-2}=1$ unless otherwise stated.

\section{Covariant perturbative approach}
\label{sec:NG_entropy}
In this section we review the evolution of primordial fluctuations on super-Hubble scales in the covariant perturbative approach. 
With the adiabatic and entropy field decomposition \cite{Langlois:2006vv} (nonlinear generalization of \cite{Gordon:2000hv}), one can clearly see how entropy modes convert into adiabatic curvature modes.

For simplicity, we consider two light, canonical scalar fields $\phi, \ \chi$ in the Friedmann-Lema\^itre-Robertson-Walker (FLRW) background:
\begin{eqnarray}
{\cal L}= \frac12\sqrt{-g}\Big[R-(\partial\phi)^2-(\partial\chi)^2-2V(\phi ,\chi)  \Big].
\end{eqnarray}
We decompose their perturbations up to second order as \cite{Langlois:2006vv}
\begin{eqnarray}
\delta\sigma &=& \cos{\theta}\delta\phi+\sin{\theta}\delta\chi+\frac{\delta s\dot{\delta s}}{2\dot\sigma},\label{eq:dsigma}\\
\delta s &=& -\sin{\theta}\delta\phi+\cos{\theta}\delta\chi -\frac{\delta\sigma}{\dot\sigma}\Big(\dot{\delta s}+\frac{\dot\theta \delta\sigma}{2}\Big),\label{eq:ds}
\end{eqnarray}
where $ \tan{\theta} \equiv \dot{\chi}/\dot\phi$ and $\dot\sigma\equiv \sqrt{\dot\phi^2+\dot\chi^2}$. Here dots denote derivatives with respect to the physical time.

The adiabatic curvature perturbation is given by 
\begin{eqnarray}\label{eq:zeta}
\zeta= -\frac{H}{\dot\sigma}\delta\sigma + \frac{H}{\dot\sigma^2}\dot\theta\delta s\delta\sigma -\frac{1}{2\dot\sigma}\left(\frac{H}{\dot\sigma}\right)^{\cdot}\delta\sigma^2,
\end{eqnarray}
which is evaluated on a flat hypersurface.
On super-Hubble scales, $\zeta$ evolves as
\begin{eqnarray}
\dot{\zeta}(t) \approx -\frac{2H}{\dot{\sigma}}\dot{\theta}\delta s+
\frac{H}{\dot{\sigma}^2}(V_{ss}+4\dot{\theta}^2){\delta s}^2 
-\frac{H}{\dot{\sigma}^3}V_{\sigma}\delta s\dot{\delta s},
\label{eq:zeta_evolution}
\end{eqnarray}
where $V_{\sigma}=\cos{\theta}V_{,\phi}+\sin{\theta}V_{,\chi}$ and $V_{ss}=(\sin{\theta})^2V_{,\phi\phi}-\sin{2\theta}V_{,\phi\chi}+(\cos{\theta})^2V_{,\chi\chi}$.

Entropy perturbations up to the second order on large scales follow
\begin{eqnarray}
\label{eq:entropy_perturbation}
\ddot{\delta s} + 3H\dot{\delta s} + (V_{ss}+3\dot{\theta}^2)\delta s \approx \hspace{3cm}\nonumber\\
-\frac{\dot{\theta}}{\dot{\sigma}}(\dot{\delta s})^2 -
\frac{2}{\dot{\sigma}}\left(\ddot{\theta}+\dot{\theta}\frac{V_{\sigma}}{\dot{\sigma}}-\frac32H\dot{\theta}\right){\delta s}\dot{\delta s}\nonumber\\
-\Big(\frac12V_{sss}-5\frac{\dot{\theta}}{\dot{\sigma}}V_{ss}-9\frac{\dot{\theta}^3}{\dot{\sigma}}\Big){\delta s}^2,
\end{eqnarray} 
where $V_{sss}=-(\sin{\theta})^3V_{,\phi\phi\phi}+3\cos{\theta}(\sin{\theta})^2V_{,\phi\phi\chi}-3(\cos{\theta})^2\sin{\theta}V_{,\phi\chi\chi}+(\cos{\theta})^3V_{,\chi\chi\chi}$. 

It is convenient to separate initial random Gaussian variables from their time evolutions as
\begin{eqnarray}
\zeta(t) &=& \zeta_* +\delta s_* {T}^{(1)}_{\zeta}(t)
+\delta s_*^2 {T}^{(2)}_{\zeta}(t),\\
\delta s(t)&=& \delta s_* {T}^{(1)}_{\delta s}(t)
+\delta s_*^2 {T}^{(2)}_{\delta s}(t),
\end{eqnarray}
where $\zeta_*$ and $\delta s_*$ are evaluated at the Hubble exit. 

One can then write the solution of Eq.~(\ref{eq:zeta_evolution}) with the transfer functions at each order:
\begin{eqnarray}
{T}^{(1)}_{\zeta}(t)   &=& \int_*^t dt'\frac{-2H}{\dot{\sigma}}\dot{\theta}{T}^{(1)}_{\delta s},\label{eq:transfer_first}\\
 {T}^{(2)}_{\zeta}(t)  &=& \int_*^t dt'\bigg[\frac{-2H}{\dot{\sigma}}\dot{\theta}{T}^{(2)}_{\delta s}+\frac{H}{\dot{\sigma}^2}(V_{ss}+4\dot{\theta}^2){{T}^{(1)}_{\delta s}}^2\label{eq:transfer_second}\nonumber\\
& & - \frac{H}{\dot{\sigma}^3}V_{\sigma}{T}^{(1)}_{\delta s}\dot{T}^{(1)}_{\delta s}\bigg].
\end{eqnarray}

The super-Hubble evolution of the power spectrum of $\zeta$ is given by
\begin{eqnarray}
\Delta_{\zeta}^2(k\ll aH)=\Bigg[\left(\frac{H}{\dot{\sigma}}\right)_*^2+\left(T_{\zeta}^{(1)}\right)^2\Bigg]\Delta_*^2 \ ,
\end{eqnarray}
where $\Delta_*^2 \equiv H_*^2/4\pi^2$.

The super-Hubble evolution of the local-form bispectrum of $\zeta$ is given by Eqs.~(\ref{eq:fnl_transfer}) and (\ref{eq:fnl_horizon}).
If $f_{NL}$ becomes large after the Hubble exit, it is important to understand the behavior of $f_{NL}^{\rm transfer}$ in detail.
In fact, as we shall see, each term in $T_{\zeta}^{(2)} [$Eq.~(\ref{eq:transfer_second})] contributes to $f_{NL}^{\rm transfer}$ equally, and thus one should not estimate it by selecting a seemingly dominant term. Otherwise, it leads to a wrong conclusion that $f_{NL}$ is large since each of them gives a large positive or negative contribution if the entropic transfer occurs.

\section{Two-field inflation and $f_{NL}$: Numerical Analysis}
\label{sec:two_field_model}

We employ a quadratic potential, such as
\begin{eqnarray}
\label{eq:two_fld_pot}
V(\phi,\chi) = \frac{m_{\phi}^2}{2}\phi^2 + \frac{m_{\chi}^2}{2}\chi^2.
\end{eqnarray} 
The inflationary trajectory of this model is characterized by a single parameter, a mass ratio $R\equiv m_{\phi}/m_{\chi}$ [see Fig.~\ref{fig:twofld_9_a}]. We assume $m_{\phi}<m_{\chi}$. Also, we investigate other potentials in Appendix~\ref{sec:appendix}.

Non-Gaussianities in two-field inflation
with the potential~(\ref{eq:two_fld_pot}) have been quantitatively studied the most thoroughly in the
literature. \citet{Vernizzi:2006ve} studied $R=1/9$ and
\citet{Yokoyama:2007dw} $R=1/20$. Both of them used the $\delta N$ formalism
and concluded that their results agree qualitatively with \cite{Rigopoulos:2005us} using the long-wavelength approach that is equivalent to the covariant approach.  
It was also shown that with various values of $R$ the $\delta N$ formalism agrees numerically very well with the covariant \cite{Watanabe:2009} or the long-wavelength approach \cite{Tzavara:2010ge}. 
\citet{Mulryne:2009kh,Mulryne:2010rp} showed that the momentum transport approach agrees with the $\delta N$ approach.

In this section, we show that the covariant perturbative approach described above agrees quantitatively with the $\delta N$ formalism for $f_{NL}$ within $1 \%$ accuracy.

 \begin{figure}
  \centering
  \includegraphics{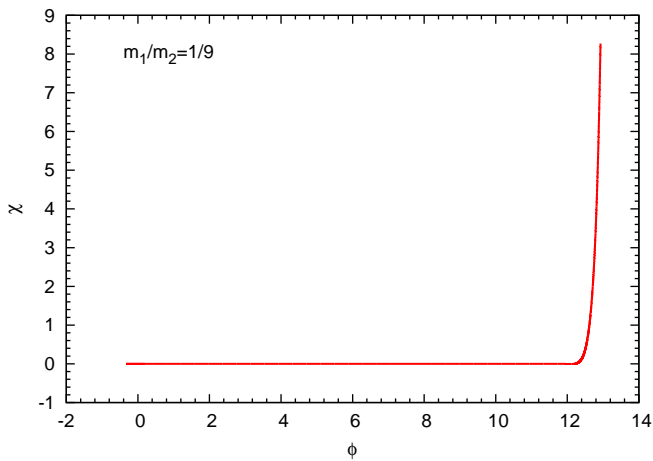} 
  \caption{A trajectory of double inflation with $R=1/9$.}
  \label{fig:twofld_9_a}
\end{figure}

\subsection{Initial conditions and background trajectories}
\label{subsec:twofld_initial_condition}
In the background the system obeys
\begin{eqnarray}
\ddot{\varphi}_I+3H\dot{\varphi}_I +V_{,\varphi^I}=0,\label{eq:klein-gordon}\\
\frac{\ddot{a}}{a}=-\frac16 (\rho+3p)=-\frac13 (\sum_I{\dot{\varphi}_I}^2-V).\label{eq:friedmann}
\end{eqnarray}
Inflationary trajectories are attractor solutions of coupled Eqs.~(\ref{eq:klein-gordon}) and
(\ref{eq:friedmann}); whatever the field values
are chosen as initial conditions, the solution flow of the system
quickly approaches to the attractor.  
We set the initial condition on the attractor by solving the equations once and restarting from the attractor solutions.

The initial conditions for perturbations are given by the linear perturbation theory \cite{Mukhanov:1990me}.
A solution of the Mukhanov-Sasaki equation in the flat gauge is given by \cite{Weinberg:2008zzc, Komatsu:2002db}
\begin{eqnarray}
\Delta^2_{\delta \varphi^I}(k)&\equiv& \frac{k^3}{2\pi^2}|\delta \varphi^I(\tau,k)|^2 = \frac{H^2}{8\pi}(-k\tau)^3|H_{\nu}^{(1)}(-k\tau)|^2\nonumber\\
&\approx& \left(\frac{H}{2\pi}\right)^2 2^{2\nu-3}\left[\frac{\Gamma(\nu)}{\Gamma(3/2)}\right]^2 \left(\frac{k}{aH}\right)^{3-2\nu},\label{eq:spectrum}
\end{eqnarray}
where $\nu^2=9/4-m^2/H^2$ and $H_{\nu}^{(1)}(x)$ is a Hankel function of the first kind. Here, $H_{\nu}^{(1)}(x\ll 1)\approx -i\Gamma(\nu)(x/2)^{-2}/\pi$ is used in the last approximation. This approximation corresponds to the super-horizon limit, $k \ll aH$. Note that $\tau\equiv \int dt/a(t)$ is the conformal time and lies in $-\infty <\tau <0$. Note also that the relation $-k\tau = k/(aH)$ is used above.

We assume light scalar fields, then $\nu\to 3/2$, for which the spectrum is independent of $k$.
We also assume that entropy modes must be light at the Hubble-horizon exit.
Therefore, our initial conditions for fluctuations are given by
\begin{eqnarray}
\Delta_{\delta \varphi^I*}^2=\Delta_{\delta s *}^2 =\Delta_{\delta \sigma *}^2= \left(\frac{H_*}{2\pi}\right)^2 \equiv \Delta_*^2,\nonumber\\
\Delta_{\zeta *}^2=\left(\frac{H_*}{\dot{\sigma}_*}\right)^2\Delta_{*}^2,\quad
\langle\delta \varphi^I_{*}({\bf k}_1) \delta \varphi^J_{*}({\bf k}_2) \delta \varphi^K_{*}({\bf k}_3)\rangle=0,\nonumber
\end{eqnarray}
where $\delta\varphi^I_*$ is evaluated in the flat gauge. The last equation implies $\delta \varphi_{*}^{I(2)}=0$, however $\delta \sigma_*^{(2)}\neq 0$ and $\delta s^{(2)}_*\neq 0$ due to Eqs.~(\ref{eq:dsigma}) and (\ref{eq:ds}).

We assume slow-roll to give initial conditions for Eqs.~(\ref{eq:zeta_evolution}) and (\ref{eq:entropy_perturbation}), and then Eqs.~(\ref{eq:zeta}) and (\ref{eq:ds}) reads \cite{Langlois:2008vk}
\begin{eqnarray}
\zeta_* &=& -\frac{1}{\sqrt{2\epsilon_*}}\delta\sigma^{(1)}_* + \frac{\eta_{ss*}}{4\epsilon_*}\delta s^2_* \nonumber\\
&&+\frac12\Big(1-\frac{\eta_{\sigma\sigma*}}{2\epsilon_*}\Big)\delta\sigma^2_*-\frac{\eta_{\sigma s*}}{2\epsilon_*}\delta\sigma_*\delta s_*,\label{eq:zeta_hr}\\
\delta s_* &=& \delta s_*^{(1)}+\frac{\eta_{ss*}}{\sqrt{2\epsilon_*}}\delta\sigma_*\delta s_* + \frac{\eta_{\sigma s*}}{2\sqrt{2\epsilon_*}}\delta\sigma^2_*,\label{eq:ds_hr}
\end{eqnarray}
where $\epsilon=\dot\sigma^2/(2H^2)$, $\eta_{\sigma\sigma}=(\cos{\theta})^2\eta_{\phi\phi}+\sin{2\theta}\eta_{\phi\chi}+(\sin{\theta})^2\eta_{\chi\chi}$, $\eta_{ss}=(\sin{\theta})^2\eta_{\phi\phi}-\sin{2\theta}\eta_{\phi\chi}+(\cos{\theta})^2\eta_{\chi\chi}$ and $\eta_{\sigma s}= \sin{2\theta}(\eta_{\chi\chi}-\eta_{\phi\phi})/2+(\cos^2{\theta}-\sin^2{\theta})\eta_{\phi\chi}$.

For the super-Hubble evolution, we calculate in two ways: (i) covariant formalism -- we solve Eqs.~(\ref{eq:zeta_evolution}) and
(\ref{eq:entropy_perturbation}) coupled with Eqs
(\ref{eq:klein-gordon}) and (\ref{eq:friedmann}) from $t=t_*$ to some
time after inflation and (ii) $\delta N$ formalism -- we solve nine adjacent background
trajectories with nine different initial field values, e.g.,
$(\phi_*+d\phi,\chi_*)$, and then compute the first and second
derivatives of the efolding number with respect to horizon crossing
field values, such as
\begin{eqnarray}\label{eq:deltaN_diff}
N_{\phi} = \frac{N(\phi+d\phi,\chi)-N(\phi-d\phi,\chi)}{2d\phi},
\end{eqnarray}
where $d\varphi^I=10^{-4}$ is chosen. We have checked that the results
converged with $10^{-4}$. For smaller values of $d\varphi^I$,
numerical noise tends to increase. 
Note that if we improve the treatment of derivatives by using the higher order of the difference than Eq.~(\ref{eq:deltaN_diff}), numerical noise would decrease but it requires more trajectories in the calculation.

For the $\delta N$ formalism, we have to set spacetime foliation as initially flat and finally uniform-Hubble (comoving) hypersurfaces.
The curvature perturbations and the power spectrum are given by \cite{Sasaki:1995aw,Sasaki:1998ug}
\begin{eqnarray}
\zeta&=&\delta N= N_{\phi}\delta\phi_*+N_{\chi}\delta\chi_* 
+\frac12N_{\phi\phi}\delta\phi_*^2 \nonumber\\
&&\hspace{2cm}+N_{\phi\chi}\delta\phi_*\delta\chi_*+\frac12N_{\chi\chi}\delta\phi_*^2,\\
\Delta_{\zeta}^2 &=& (N_{\phi}^2+N_{\chi}^2)\Delta_*^2.
\end{eqnarray}
 Note that Eqs.~(\ref{eq:zeta_evolution}) and (\ref{eq:entropy_perturbation}) are gauge-invariant \cite{Langlois:2006vv}, thus we set the same foliation for a comparison purpose.

\begin{figure}
  \centering\vspace{-2.7cm}
  \includegraphics{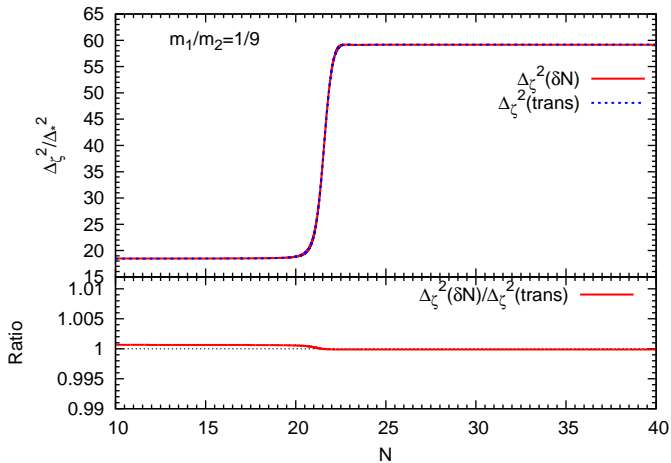}
  \caption{Evolution of the power spectrum of $\zeta$, which is normalized by $\Delta_*^2\equiv H_*^2/(4\pi^2)$. Here $\Delta_{\zeta}^2(\delta N)$ indicates the $\delta N$ formalism, while $\Delta_{\zeta}^2({\rm trans})$ indicates the covariant formalism.}
  \label{fig:twofld_9_aa}
\end{figure}

\begin{figure}
  \centering\vspace{-2cm}
  \includegraphics{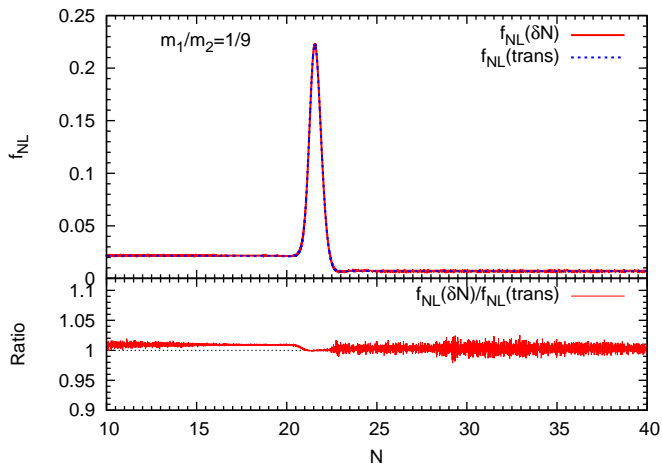}
  \caption{Evolution of non-Gaussianity, $f_{NL}$. Here $f_{NL}(\delta N)$ indicates the $\delta N$ formalism, while $f_{NL}({\rm trans})$ indicates the covariant formalism.}
  \label{fig:twofld_9_ba}
\end{figure}

\begin{figure}
  \centering
  \includegraphics{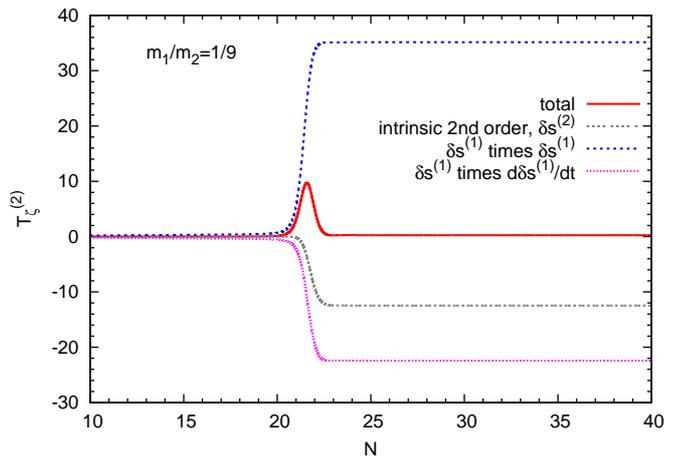}	
  \caption{Evolution of $\zeta^{(2)}$. The red line shows Eq.~(\ref{eq:transfer_second}). The gray dash-dotted, blue dashed and pink dotted lines show the first, second and third terms in Eq.~(\ref{eq:transfer_second}), respectively.}
  \label{fig:twofld_9_b}
\end{figure}

\subsubsection{double inflation with $m_{\phi}/m_{\chi}=1/9$}
We set the same initial condition as \cite{Vernizzi:2006ve}.  Initial
field values are $\phi_i=\chi_i=13$. We assume that inflation ends at
$\epsilon_e=1$ and the number of efolds at the end of inflation from
the Hubble-horizon exit is $N_e=60$.  We find that $(\phi_*=12.9281,
\chi_*=8.27202)$. The perturbed equations are integrated from this
time.

Figure~\ref{fig:twofld_9_a} shows the trajectory in the
field space, where one can find a \textit{turn}. At the turn,
entropy perturbations source $\zeta$ through
Eq.~(\ref{eq:zeta_evolution}). Since the entropy mode becomes heavy
and decays sufficiently after the turn, it no longer sources
$\zeta$. As shown in Fig.~\ref{fig:twofld_9_aa} and
Fig.~\ref{fig:twofld_9_ba}, both amplitude and non-Gaussianity of $\zeta$ are
generated during the turn. Both figures show agreement of two formalisms with $\lesssim 1\%$ accuracy. The
hight of the peak is $\sim 0.22$ which agrees precisely in both the
$\delta N$ formalism and the one calculated from
Eqs.~(\ref{eq:fnl_transfer}) and (\ref{eq:fnl_horizon}). Note that
$f_{NL}$ computed here corresponds to $f_{NL}^{(4)}$ in
\cite{Vernizzi:2006ve}.
Since the evolution of $f_{NL}^{(3)}$ in \cite{Vernizzi:2006ve} is governed only by $N_I$ and $T_{\zeta}^{(1)}$ for each formalism \cite{RenauxPetel:2009sj}, the agreement is the same order as  the power spectrum $\Delta_{\zeta}^2$.

How can we understand the feature of non-Gaussianity appeared during the
turn? The cause of the tiny net effect is the precise cancellation
between terms in Eq.~(\ref{eq:zeta_evolution}). 
As mentioned in Sec.~\ref{sec:NG_entropy}, the shape of $f_{NL}$ is attributed to that of the second order transfer function $T_{\zeta}^{(2)}$ [Eq.~(\ref{eq:transfer_second})]. As shown in Fig.~\ref{fig:twofld_9_b}, the second term has the sharpest rise; and then negative contributions from the first and the third terms catch up with it. The value after the turn is given by $f_{NL}\sim 0.007$, which consists of $f_{NL}^{\rm
transfer}\sim 0.005$ and $f_{NL}^{\rm horizon}\sim 0.002$. The value
before the turn comes completely from $f_{NL}^{\rm horizon}\sim 0.021$.

\begin{figure}
  \centering
  \includegraphics{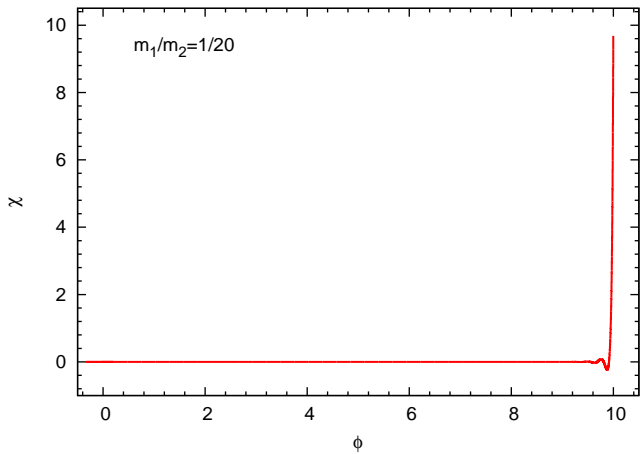}
  \caption{A trajectory of double inflation with $R=1/20$.}
  \label{fig:twofld_20_a}
\end{figure}

\begin{figure}
  \centering\vspace{-2.5cm}
  \includegraphics{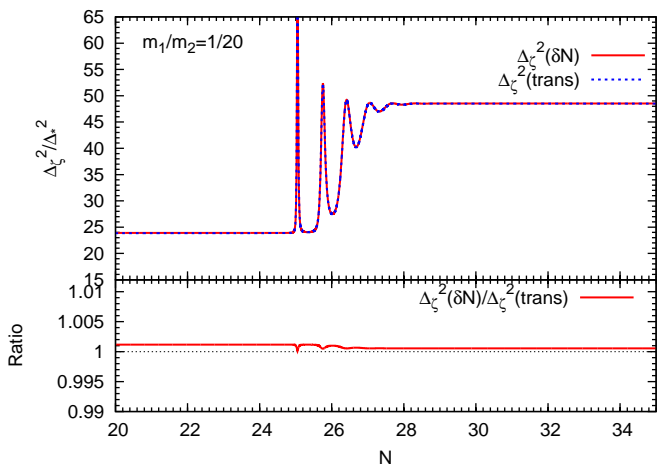}
  \caption{Evolution of the power spectrum of $\zeta$, which is normalized by $\Delta_*^2\equiv H_*^2/(4\pi^2)$.}
  \label{fig:twofld_20_aa}
\end{figure}

\begin{figure}
  \centering\vspace{-2.5cm}
  \includegraphics{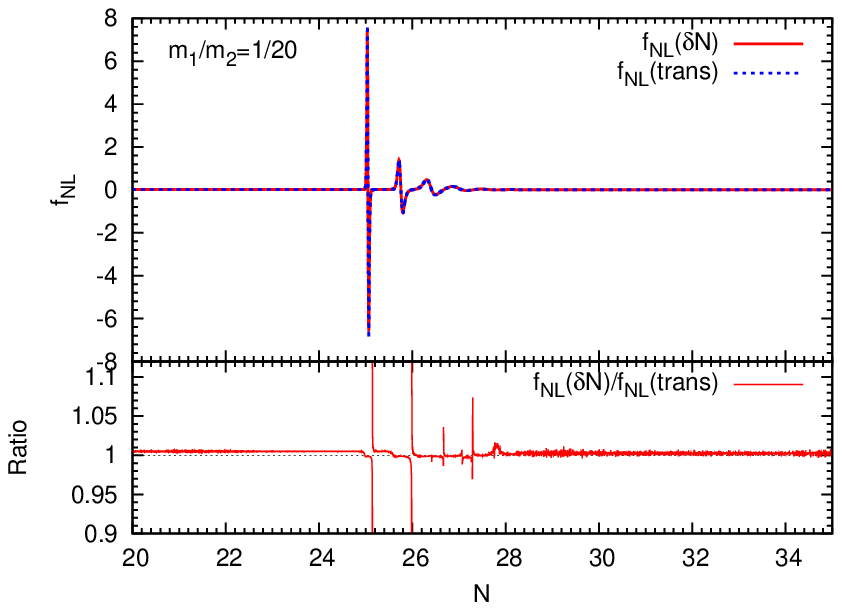}
  \caption{Evolution of non-Gaussianity, $f_{NL}$. The ratio is divergent (ill-defined) when $f_{NL}$(trans) crosses zero.}
  \label{fig:twofld_20_ba}
\end{figure}

\begin{figure}
  \centering
\includegraphics{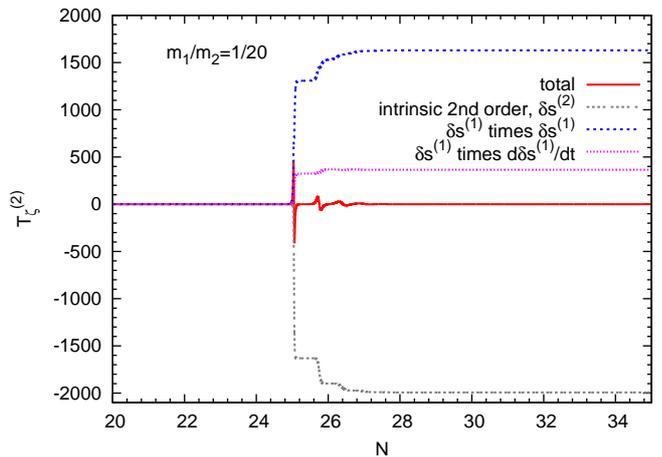}
  \caption{Evolution of $\zeta^{(2)}$.}
  \label{fig:twofld_20_b}
\end{figure}

\subsubsection{double inflation with $m_{\phi}/m_{\chi}=1/20$}
We set a similar initial condition to \cite{Yokoyama:2007dw}. Initial field values are set to $\phi_i=\chi_i=10$, and we find $(\phi_*=9.99920 ,\chi_*= 9.68230 )$ by setting $N_e=50$.

The similar feature as the case of $R=1/9$ can be found in
Figs.~\ref{fig:twofld_20_a}, \ref{fig:twofld_20_aa}, and \ref{fig:twofld_20_ba}. There are a
few sub-turns in a turn. Each sub-turn creates a series of peak
and dip.  Since the mass ratio is larger, the turn is sharper and
leaves features on the power spectrum and bispectrum. Again, the $\delta N$ and covariant formalisms agree with $\lesssim 1\%$ accuracy. The
peak value is $f_{NL}\sim 7.5$. The value after the turn is
$f_{NL}\sim 0.001$. The value before the turn is $f_{NL}\sim 0.002$.
One can clearly see the very delicate cancellation among the entropic transfer terms in Fig.~\ref{fig:twofld_20_b}.

From the above two cases, one can gain some insights of the mechanism to
generate non-Gaussianity at the turn in the field space. The sharper the turn is, the more non-Gaussianity is peaked. A peak is followed by a dip due to different timing contributions from each term in Eq.~(\ref{eq:transfer_second}); the net effect erases almost all the traces due to the attractor behavior after the turn. The residual non-Gaussianity can remain if the turn is slow enough to allow the background quantities, such as the expansion rate, change their values during the turn before entropy modes decay.

We have checked that the accurate agreement also holds for other parameters and other potentials, e.g., $V=m_{\chi}^2\chi^2e^{-\lambda\phi^2}/2$ used in \cite{Byrnes:2008wi,Mulryne:2009kh} and $V= a_2\chi^2+b_0-b_2\phi^2+b_4\phi^4$ used in \cite{Tzavara:2010ge} (see Appendix~\ref{sec:appendix}).

\section{Analytic estimate of damping $f_{NL}$}

In the previous section we have shown that how the cancellation of $T^{(2)}_{\zeta}$ happens numerically when the entropy mode damps, i.e., the inflationary trajectory becomes purely adiabatic since the entropy direction closes.

In this section, we analyze an envelope behavior of $f_{NL}$ when the trajectory approaches the adiabatic limit by solving the perturbed equations while leaving the form of the potential arbitrary.
In this case, there are two regimes. Initially, entropy modes are light and source adiabatic modes. Eventually, entropy modes get heavier and damp away. We solve the super-Hubble evolution of $\zeta$ in the latter regime by dividing cases [A for overdamped (light) entropy modes, B and C for underdamped (heavy) entropy modes], and we will find how fast $f_{NL}$ approaches to the asymptotic value that can be large in principle.

Physically, case A describes the slow-roll regime of two-field inflation, where curvature perturbations have already been sourced by the time $t_1 > t_*$ but entropy modes are still light. An example of this is double inflation with $R$ close to 1.
Cases B and C describe the slow-roll regime, where entropy modes get heavy after sourcing $\zeta$. Examples of these are double inflation with $R=1/4$ for B and $R=1/20$ for C. These three cases cover almost all two-field models when entropy modes decay while separating cases in actual models is subtle.

\subsection{Small $\eta_{ss}$ and small $\dot\theta$}
When entropy modes are light ($\eta_{ss} < 3/4$), one may use slow-roll equations for entropy modes to first order:
\begin{eqnarray}
\dot{\delta s} = - \eta_{ss}H\delta s,\label{eq:ds_sr}
\end{eqnarray}
where $\dot{\delta s}$ have been used in deriving the initial conditions $\zeta_*, \ \delta s_*$ in the previous section.
We integrate these equations further by assuming some entropic transfer happened by the time $t_1>t_*$.

The solution of Eq.~(\ref{eq:ds_sr}) is given by
\begin{eqnarray}
\delta s &=& \delta s_1 \exp \Big[-\int_{t_1} dt H \eta_{ss}\Big] \nonumber\\ 
&\simeq& \delta s_1 \left(\frac{a}{a_1}\right)^{-\eta_{ss}},\quad
\eta_{ss}  \simeq {\rm const.},\label{eq:ds_sr_sol}
\end{eqnarray}
where we have used a slow roll relation $\dot{\eta}_{ss}=2H\epsilon\eta_{ss}+2H\eta_{\sigma s}^2-H\xi_{\sigma ss}^2 \simeq 0$ \cite{Byrnes:2006fr}\footnote{This solution can be also obtained from Eq.~(\ref{eq:spectrum}) by setting $m^2/H^2 = 3\eta_{ss}$ and expanding for small $\eta_{ss}$ as $\nu \simeq 3/2-\eta_{ss}$.}. 

The super-Hubble evolution of $\zeta$ is  then given by
\begin{eqnarray}
\dot\zeta&=&-\frac{2H}{\dot\sigma}\dot\theta\delta s +\frac{4H}{\dot\sigma^2}\dot\theta^2\delta s^2 \nonumber\\
 &=& \sqrt{\frac{2}{\epsilon}}\eta_{\sigma s}H\delta s+\frac{2\eta_{\sigma s}^2}{\epsilon}H\delta s^2,
\end{eqnarray}
where we have used slow roll relations: $3H\dot\sigma\simeq -V_{\sigma}$, $\dot{\delta s}\simeq -\eta_{ss}H\delta s$ and $\dot\theta/H \simeq -\eta_{\sigma s} \simeq {\rm constant}$.
Using Eq.~(\ref{eq:ds_sr_sol}), one can integrate $\zeta$ as
\begin{eqnarray}
\zeta \simeq \zeta_1  -\frac{\eta_{\sigma s}}{\eta_{ss}}\sqrt{\frac{2}{\epsilon}}\delta s_1\left(\frac{a}{a_1}\right)^{-\eta_{ss}}
-\frac{\eta_{\sigma s}^2}{\epsilon \eta_{ss}}\delta s^2_1 \left(\frac{a}{a_1}\right)^{-2\eta_{ss}},
\end{eqnarray} 
where $\zeta_1$ is the asymptotic value that is different from Eq.~(\ref{eq:zeta_hr}) in general. The second order term should cancel out with the time evolution of $\delta s^{(2)}$, but we did not catch the feature in this estimation; we just obtained that the second order term damps away eventually if $\eta_{ss}$ is sizable. In this case, $f_{NL}\sim \zeta^{(2)}/\zeta_1^2$ damps as $a^{-2\eta_{ss}}$ toward its asymptotic value.

\subsection{Large $\eta_{ss}$ and small $\dot\theta$}
In this case one cannot use the slow roll relations for entropy modes but can do along the trajectory, i.e., in $\sigma$-direction.

From Eq.~(\ref{eq:spectrum}) one has on the super-Hubble scale
\begin{eqnarray}
|\delta s^{(1)}| &=& \frac{2^{{\rm Re}(\nu)-3/2}}{\sqrt{2H}}\frac{|\Gamma(\nu)|}{\Gamma(3/2)}a^{-3/2}\left(\frac{k}{aH}\right)^{-{\rm Re}(\nu)},\nonumber\\
\nu &=&  \sqrt{\frac94 - \Big(3\eta_{ss}+\frac{3\dot\theta^2}{H^2}\Big)},\label{eq:ds_heavy}
\end{eqnarray}
where $\nu$ is pure imaginary if $\eta_{ss}> 3/4$ and/or $\dot\theta^2/H^2 > 3/4$.
Here, we shall assume large $\eta_{ss}> 3/4$ while keeping other slow roll parameters small. This assumption leads to slow roll and slow turn, $\dot\theta \simeq -(1+\eta_{ss}/3)\eta_{\sigma s}H$.
In this case, we have $\delta s^{(1)}\propto a^{-3/2}$. 

The super-Hubble evolution of $\zeta$ is given by
\begin{eqnarray}\label{eq:zeta_srst}
\dot\zeta = -\frac{2H }{\dot\sigma}\dot\theta\delta s 
+\frac{ H}{\dot\sigma^2}(V_{ss}+4\dot\theta^2-\frac92 H^2)\delta s^2,
\end{eqnarray}
where we have used $3H\dot\sigma\simeq -V_{\sigma}$ and $\dot{\delta s}=-(3/2)H\delta s$. 
Since $\dot\eta_{\sigma s}\simeq 2H\epsilon\eta_{\sigma s}+H\eta_{\sigma s}(\eta_{\sigma\sigma}-\eta_{ss})-\xi_{\sigma\sigma s}^2 \simeq -H\eta_{\sigma s}\eta_{ss}$ for $\eta_{ss}\gg 1$, we get
\begin{eqnarray}
\eta_{\sigma s} = \eta_{\sigma s1} \exp \Big[-\int_{t_1}dt H \eta_{ss}\Big] \simeq \eta_{\sigma s1} \left(\frac{a}{a_1} \right)^{-\eta_{ss}}.
\end{eqnarray}

\begin{figure}
  \centering
  \includegraphics{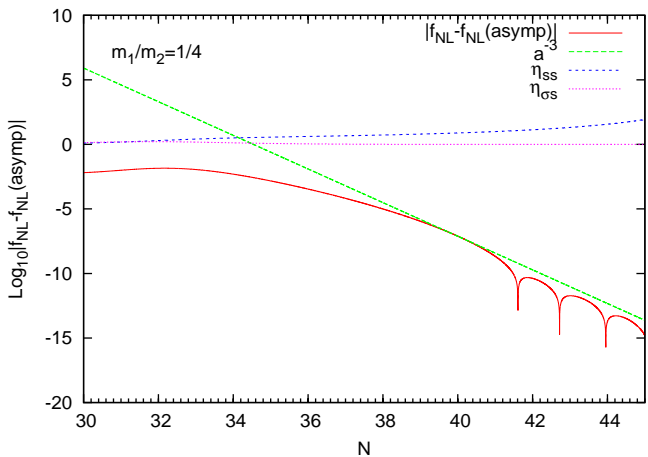}
  \caption{Damping of non-Gaussianity, $\log_{10}|f_{NL}-f_{NL}^{\rm asymp}|$ in the case of $R=1/4$, where $f_{NL}^{\rm asymp}= 0.00855163$. The envelope scales as $a^{-3}$. The values of $\eta_{ss}
  $ and $\eta_{\sigma s}$ in a linear scale are also shown for reference.}
  \label{fig:fnl_4_fate}
\end{figure}

Using this and Eq.~(\ref{eq:ds_heavy}), one can integrate $\zeta$ as
\begin{eqnarray}\label{eq:zeta_heavy}
\zeta &\simeq& \zeta_1-\frac{ \eta_{\sigma s1}(\eta_{ss}/3+1) }{(\eta_{ss}+3/2)}\sqrt{\frac{2}{\epsilon}}\delta s_1 \left(\frac{a}{a_1}\right)^{-\eta_{ss}-3/2} \nonumber\\
&&-\frac{\eta_{ss}-3/2 }{2\epsilon}\delta s_1^2 \left(\frac{a}{a_1}\right)^{-3}\nonumber\\
&&-\frac{ \eta_{\sigma s1}^2(\eta_{ss}/3+1)^2}{(\eta_{ss}+3/2)\epsilon} \delta s_1^2 \left(\frac{a}{a_1}\right)^{-2\eta_{ss}-3} 
,
\end{eqnarray}
where we have assumed constancy of $\eta_{ss}$. 
For $\eta_{ss}> 3/4$, the value of $f_{NL}$ decays as $a^{-3}$ since the second line of Eq.~(\ref{eq:zeta_heavy}) dominates [see fig.~\ref{fig:fnl_4_fate}].

\citet{Meyers:2010rg} have recently claimed that $f_{NL}$ decays exponentially when the entropy mode damps by using the $\delta N$ formalism (see Eq.~(72) in \cite{Meyers:2010rg}):
\begin{eqnarray}
f_{NL} \sim {\cal O}(\epsilon_*) + {\cal O}(1)\times \eta_{ss} \exp\Big[{-2\int_{t_1}dt H C_{\eta}\eta_{ss}}\Big],
\end{eqnarray}
where $C_{\eta}$ is a number greater than 1. This formula would correspond to the last term of Eq.~(\ref{eq:zeta_heavy}) if $\eta_{ss}\sim {\rm const.}\gg 3/2$. 
This exponential behavior is valid only in the last tail of $f_{NL}$ that approaches to its asymptotic value and shows up after the oscillatory feature of entropy modes disappears.

\subsection{Large $\eta_{ss}$ and large $\dot\theta$}
\begin{figure}
  \centering
  \includegraphics{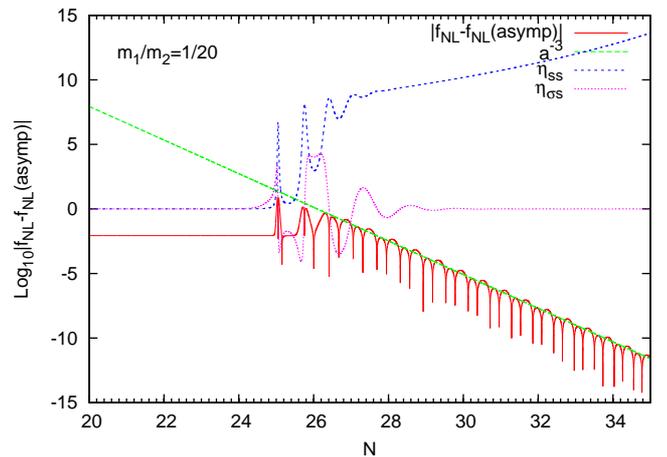}
  \caption{Damping of non-Gaussianity, $\log_{10}|f_{NL}-f_{NL}^{\rm asymp}|$ in the case of $R=1/20$, where $f_{NL}^{\rm asymp}=0.00854551$. The envelope scales as $a^{-3}$.}
  \label{fig:fnl_20_fate}
\end{figure}

If the angle changes rapidly, one cannot use the slow roll relations for entropy modes or slow turn relations but still can do in $\sigma$-direction. Instead, we use the scaling of $\dot\theta =(-\dot\phi V_{,\chi}+\dot\chi V_{,\phi})/\dot\sigma \sim a^{-3/2}$ since the heavy field decays as $a^{-3/2}$ (if $\phi$ is heavy, $\dot\phi \sim a^{-3/2}$, $V_{,\phi}\sim a^{-3/2}$ or faster depending on models). In this case, we can integrate Eq.~(\ref{eq:zeta_srst}) as 
\begin{eqnarray}
\zeta \simeq \zeta_1 -\frac{ C_{\theta} }{3}\sqrt{\frac{2}{\epsilon}}\delta s_1 \left(\frac{a}{a_1}\right)^{-3} \hspace{3cm}\nonumber\\
-\frac{\eta_{ss}-3/2 }{2\epsilon}\delta s_1^2 \left(\frac{a}{a_1}\right)^{-3}
-\frac{C_{\theta}^2}{3\epsilon} \delta s_1^2 \left(\frac{a}{a_1}\right)^{-6}, 
\end{eqnarray}
where $-\dot\theta/H \sim C_{\theta}(a/a_1)^{-3/2}$ have been used.
Thus, we obtain $f_{NL}\sim \eta_{ss}(a/a_1)^{-3}$ if $C_{\theta}\sim {\cal O}(1)$ [see fig.~\ref{fig:fnl_20_fate}].

\section{Conclusion}
\label{sec:conclusion_NG}
We have re-examined the super-Hubble evolution of the primordial non-Gaussianity in two-field
inflation by taking two approaches: the $\delta N$ and the covariant perturbative formalisms. 
The results agree within $1\%$ accuracy in the cases of two-field
inflation with the quadratic potential~(\ref{eq:two_fld_pot}) and with the other potentials in Appendix~\ref{sec:appendix}.  
The $\delta N$ formalism and the long wavelength formalism, which is equivalent to the covariant one, were also compared analytically and numerically in \cite{Tzavara:2010ge}, and our results agree with theirs.
Our treatment has improved the agreement in the previous literature.

The peak feature appears on $f_{NL}$ at the turn in the field space, which can be understood as
the precise cancellation between terms in Eq.~(\ref{eq:transfer_second}).  
Note, however, that we have made
the slow-roll approximation for both approaches at the horizon exit even though integration after the horizon exit has been carried out without assuming slow roll.

In Figs.~\ref{fig:twofld_9_b} and \ref{fig:twofld_20_b}, we have shown that the net effect of the second order entropic transfer is relatively small due to the cancellation between terms in Eq.~(\ref{eq:transfer_second}) with the quadratic potential, which also holds for the other potentials studied in Appendix~\ref{sec:appendix} if the evolution is tracked until entropy modes decay.
Thus, it is important to include all terms in the second order perturbation theory for correctly estimating $f_{NL}$. 
If $f_{NL}$ is detected by observations, the two-field model with Eq.~(\ref{eq:two_fld_pot}) is ruled out. 

Our method to calculate the second order $\zeta$ can thus be used for cross-checking results with two formalisms, where careful evaluations of Eq.~(\ref{eq:transfer_second}) are necessary. In addition, inclusion of Eq.~(\ref{eq:fnl_horizon}) is crucial to obtain the accurate agreement.

We have analytically shown that if large $f_{NL}$ is generated during inflation, it decays as $a^{-3}$ approaching to the adiabatic limit in a broad class of two-field models. In principle, persistent values of non-Gaussianity, $f_{NL}^{\rm asymp}$, may be observably large, but we did not meet such an example. 

Finally, let us mention the relative merits of the covariant perturbative approach. This approach requires a single
realization in the simulation; statistical properties are taken into
account in the formulation. It will significantly save a lot of computational
time if integration of the background equations is time consuming since the $\delta N$ formalism requires many realizations, especially if one needs to deal with many scalar fields. It is straightforward to generalize our formulation to N-fields. Also, the numerical noise is much less for the covariant formalism than the $\delta N$ formalism. 
It still remains to be seen a computation of non-Gaussianity by the covariant perturbative approach coupled with a lattice simulation that is suitable for preheating.

\begin{acknowledgments}
The author wishes to thank Chris Byrnes, Angela Lepidi, David Mulryne and Filippo Vernizzi for useful comments on the draft. He thanks Donghui Jeong for discussions and advices on numerical calculations and Eiichiro Komatsu for many discussions and comments. He would also like to thank the ICG, Portsmouth, especially Kazuya Koyama, for the invitation to the informal workshop and their kind hospitality, when this work was at the final stage. This work is supported by the TRR 33 ``The Dark Universe". 
\end{acknowledgments}

\appendix

\begin{figure}
\centering
\includegraphics{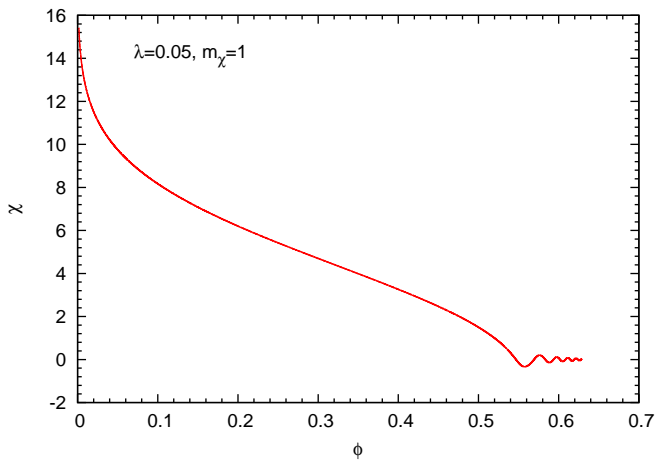}
\caption{A trajectory of inflation model with Eq.~(\ref{eq:exp_pot}).}
\label{fig:fldspace_exp}
\end{figure}

\begin{figure}
\centering\vspace{-2.7cm}
\includegraphics{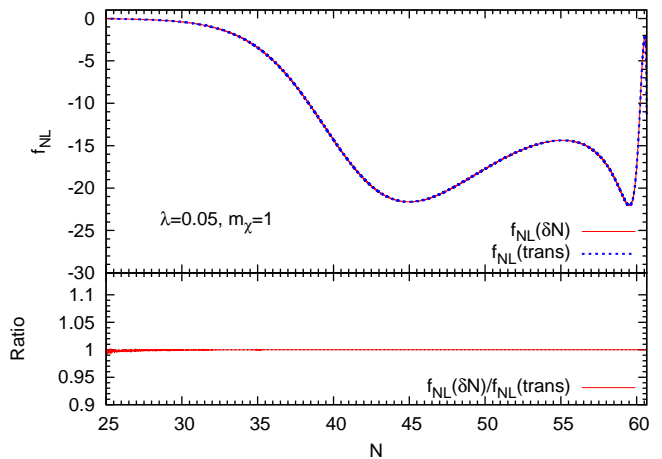}
\caption{Evolution of non-Gaussianity, $f_{NL}$ for Eq.~(\ref{eq:exp_pot}).}
\label{fig:fnl_exp_e}
\end{figure}

\section{Other examples: ``large" non-Gaussianity}\label{sec:appendix}
In order to back up the results obtained, we further study two illuminating models which produce relatively ``large" non-Gaussianities, $|f_{NL}|\sim {\cal O}(1)$, in the sense that the values are much larger than those from single-field inflation and two-field inflation with Eq.~(\ref{eq:two_fld_pot}).
However, since the upcoming Planck satellite is expected to constrain $\Delta f_{NL}\sim {\cal O}(5)$ \cite{Komatsu:2001rj}, these models do not predict $f_{NL}$ large enough to be observed. Nevertheless, they are interesting examples.

The models are    
\begin{eqnarray}
V(\phi,\chi)&=&\frac{m_{\chi}^2}{2}\chi^2e^{-\lambda\phi^2},\label{eq:exp_pot}\\
V(\phi,\chi)&=&a_2\chi^2+b_0-b_2\phi^2+b_4\phi^4,\label{eq:asym_pot}
\end{eqnarray}
which have been studied in \cite{Byrnes:2008wi,Mulryne:2009kh} and in \cite{Tzavara:2010ge}, respectively.

For Eq.~(\ref{eq:exp_pot}), we set a similar initial condition to \cite{Byrnes:2008wi,Mulryne:2009kh}. Initial field values are set to $(\phi_i =10^{-3} ,\  \chi_i =16 )$, and we find $(\phi_*= 0.00155720,\ \chi_*=15.4210)$ by setting $N_e = 60$. The model parameters are chosen to be $\lambda = 0.05,\ m_{\chi}=1$ in numerical units to match with \cite{Byrnes:2008wi,Mulryne:2009kh}.

Figs.~\ref{fig:fldspace_exp} and \ref{fig:fnl_exp_e} show that there are smooth, continuous turns generating non-Gaussianity of $\zeta$ until oscillations in $\chi$-direction start. Inflation ends at $\epsilon=1$ and $N=60$, which is right before the oscillation phase.
The generated $f_{NL}$ is negative and large, but transient. After the entropy mode gets heavy, the asymptotic value of $f_{NL}$ stays around $f_{NL}^{\rm asymp}\sim -2.1$ and a series of peak and dip shows up at each oscillation in $\chi$.
Since each series of peak and dip at each oscillation cancels out precisely, $f_{NL}$ approaches to its asymptotic value as $a^{-3}$.
This fact was not pointed out previously in \cite{Mulryne:2009kh}; they stopped solving the evolution equations at the end of inflation.

\begin{figure}[t]
\centering
\includegraphics{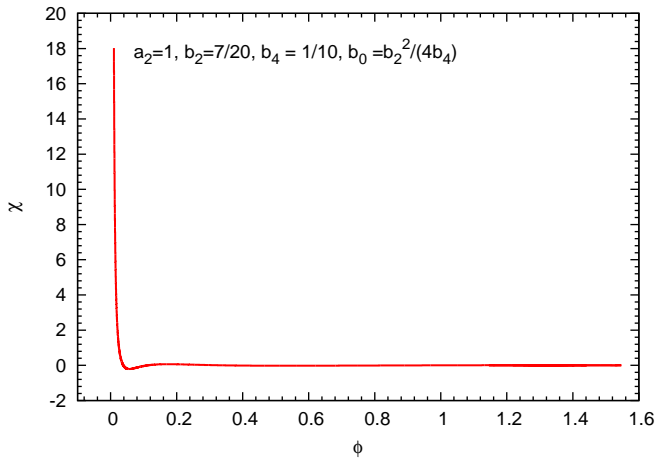}
\caption{A trajectory of inflation model with Eq.~(\ref{eq:asym_pot}).}
\label{fig:fldspace_asym}
\end{figure}

For Eq.~(\ref{eq:asym_pot}), we set a similar initial condition to \cite{Tzavara:2010ge}. Initial field values are set to $(\phi_i = 0.01,\ \chi_i=18.1)$, and we find $(\phi_*=0.0100143,\ \chi_*=18.0239)$ by setting $N_e=85$. The model parameters are chosen to be $a_2=1,\ b_2=7/20,\ b_4=1/10, b_0=b_2^2/(4b_4)$ in numerical units to match with \cite{Tzavara:2010ge}.

\begin{figure}[t]
\centering\vspace{-2.5cm}
\includegraphics{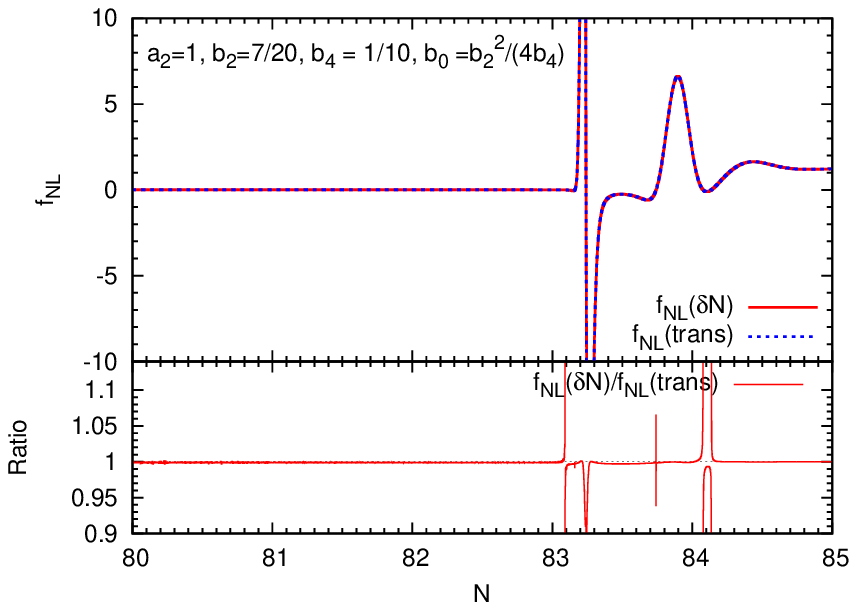}
\caption{Evolution of non-Gaussianity, $f_{NL}$ for Eq.~(\ref{eq:asym_pot}). The ratio is divergent (ill-defined) when $f_{NL}$(trans) crosses zero.}
\label{fig:fnl_asym_e}
\end{figure}

Figs.~\ref{fig:fldspace_asym} and \ref{fig:fnl_asym_e} show that there are a few turns generating a few series of peak and dip in $f_{NL}$. The first peak value is as large as $f_{NL}\sim 40$. As in the case of Eq.~(\ref{eq:exp_pot}), $f_{NL}$ approaches to a persistent value of $f_{NL}^{\rm asymp}\sim 1.224$ as $a^{-3}$.

\citet{Tzavara:2010ge} have also compared the $\delta N$ formalism and the long wavelength formalism, which is in principle equivalent to the covariant formalism after choosing a coordinate system and expanding to the lowest order in spatial gradients.
In the case of Eq.~(\ref{eq:asym_pot}), they have reported asymptotic values of $f_{NL}^{\rm LWF}\sim 1.19$ for the long wavelength formalism and $f_{NL}^{\delta N}\sim 1.23$ for the $\delta N$ formalism; they agree within 3.6\%. Our formula for $f_{NL}$ [Eqs.~(\ref{eq:fnl_transfer}) and (\ref{eq:fnl_horizon})] corresponds to Eq.~(3.6) in \cite{Tzavara:2010ge}. In there, $\bar{v}_{12}$ corresponds to our $\sqrt{2\epsilon_*}T_{\zeta}^{(1)}$, $g_{int}+g_{iso}$ to our $-2\epsilon_* T_{\zeta}^{(2)}$, $\tilde{W}_{22}$ to our $\eta_{ss}$, $\epsilon-\eta^{\|}$ to our $\eta_{\sigma\sigma}$, and $-\eta^{\bot}$ to our $\eta_{\sigma s}$ in slow roll.

%\pagebreak

\bibliographystyle{apsrev}
\bibliography{watanabe}

\end{document}